\documentclass[lettersize,journal]{IEEEtran}
\usepackage{amsmath,amsfonts}
\usepackage{algorithmic}
\usepackage{array}
\usepackage[caption=false,font=normalsize,labelfont=sf,textfont=sf]{subfig}
\usepackage{textcomp}
\usepackage{stfloats}
\usepackage{url}
\usepackage{verbatim}
\usepackage{graphicx}
\def\BibTeX{{\rm B\kern-.05em{\sc i\kern-.025em b}\kern-.08em
    T\kern-.1667em\lower.7ex\hbox{E}\kern-.125emX}}
\usepackage{balance}

\usepackage{orcidlink}
\usepackage{kotex}
\usepackage{mathptmx}
\usepackage{booktabs}
\usepackage{multirow}
\usepackage{amsmath}
\usepackage{xcolor}

\graphicspath{{figs/}{figures/}{pictures/}{images/}{./}}

\begin{document}

\title{d-DQIVAR: Data-centric Visual Analytics and Reasoning for Data Quality Improvement}

\author{Hyein Hong\orcidlink{0000-0003-1410-1788}, Sangbong Yoo\orcidlink{0000-0002-0973-9288},~\IEEEmembership{Member,~IEEE}, SeokHwan Choi\orcidlink{0009-0009-6452-3234}, Jisue Kim\orcidlink{0009-0002-6643-5835}, Seongbum Seo\orcidlink{0000-0002-9582-1674}, Haneol Cho, Chansoo Kim, and Yun Jang\orcidlink{0000-0001-7745-1158},~\IEEEmembership{Member,~IEEE}
\thanks{H. Hong, S. Choi, S. Seo, H. Cho, and Y. Jang are with Sejong University, Seoul, South Korea. E-mail: \{gumdung98\,$|$\,my2853121\}@naver.com, seo@seongbum.com, haneol815@gmail.com, jangy@sejong.edu}
\thanks{S. Yoo and C. Kim are with AI, Information and Reasoning (AI/R) Laboratory, Korea Institute of Science and Technology (KIST), Seoul, South Korea. E-mail: \{usangbong\,$|$\,eau\}@kist.re.kr}
\thanks{J. Kim is with Wavebridge, South Korea, E-mail: jisue0216@gmail.com}
\thanks{H. Hong and S. Yoo contributed equally to this work.}
\thanks{Y. Jang is the corresponding author.}}

\markboth{IEEE Transactions on Visualization and Computer Graphics,~Vol.~xx, No.~xx, xx~20xx}%
{Hong \MakeLowercase{\textit{et al.}}: VARIUS: Visual Analytics and Reasoning for Informed User-Guided Selection of Data Quality Procedures in Machine Learning Tasks}


\IEEEtitleabstractindextext{%
    \begin{abstract}
        Approaches to enhancing data quality~(DQ) are classified into two main categories: data- and process-driven. However, prior research has predominantly utilized batch data preprocessing within the data-driven framework, which often proves insufficient for optimizing machine learning~(ML) model performance and frequently leads to distortions in data characteristics. 
        Existing studies have primarily focused on data preprocessing rather than genuine data quality improvement~(DQI). In this paper, we introduce d-DQIVAR, a novel visual analytics system designed to facilitate DQI strategies aimed at improving ML model performance. Our system integrates visual analytics techniques that leverage both data-driven and process-driven approaches. 
        Data-driven techniques tackle DQ issues such as imputation, outlier detection, deletion, format standardization, removal of duplicate records, and feature selection. Process-driven strategies encompass evaluating DQ and DQI procedures by considering DQ dimensions and ML model performance and applying the Kolmogorov-Smirnov test. We illustrate how our system empowers users to harness expert and domain knowledge effectively within a practical workflow through case studies, evaluations, and user studies.
    \end{abstract}
    
}

\maketitle
\IEEEdisplaynontitleabstractindextext
\section{Introduction}
\IEEEPARstart{T}{he} efficacy of Machine Learning (ML) models is significantly impacted by Data Quality (DQ)~\cite{atla2011sensitivity}. Various strategies, both data-driven and process-driven, are used in Data Quality Improvement (DQI) methods to bolster ML model performance~\cite{batini2009methodologies}. However, prevailing research has predominantly focused on data preprocessing methods aligned with data-driven strategies. Consequently, users often redundantly compare ML model performance based on data alterations without integrating process-driven strategies. Furthermore, prevalent research relies on batch preprocessing techniques to handle common DQ issues such as missing values or outliers. Nevertheless, the optimal DQI approach may vary depending on specific objectives~\cite{li2019cleanml}. Hence, this approach might not be optimal for enhancing ML model performance.

Recently, a DQI approach was introduced to enhance ML performance. This approach integrates data- and process-driven strategies within a visual analytics (VA) system~\cite{hong2023how}. However, its practical usability has faced criticism. Users are expected to possess substantial expertise, often encountering information overload and implicit visualizations during the process. The system exhibited limitations in its adaptability, particularly in tailoring DQI to specific data domains or user requirements, resulting in reduced flexibility and usability. These limitations arose because the system primarily focused on providing a singular DQI process to address past challenges. Although large language models have been employed to minimize expert intervention, their limited integration of human input poses challenges in tailoring DQI to specific task contexts~\cite{wei2025zeroed}. Since contextual adaptation still relies on expert input, visual reasoning offers a practical solution for tailoring DQI to specific tasks.

In this paper, we present d-DQIVAR, a novel data-centric visual analytics system to improve DQ in ML tasks. d-DQIVAR utilizes VA techniques that leverage data-driven and process-driven strategies to address DQ issues typically encountered in conventional data preprocessing. As part of our data-driven strategy, we offer comprehensive DQI methods necessary for enhancing ML model performance. These methods include imputation, outlier detection and removal, format standardization, removal of duplicate records, and feature selection. Additionally, we incorporate a process-driven strategy involving DQ dimensions, ML model performance evaluation, and the Kolmogorov-Smirnov (K-S) test to assist users in assessing DQ and DQI procedures and determining the most suitable DQI approach based on these assessments. d-DQIVAR enhances usability by supporting visual reasoning for decision-making, integrating conventional visualizations and statistical preprocessing to provide comprehensive insights into DQI. We illustrate the usage scenarios of the system through case studies conducted on two datasets. Furthermore, we validate the practicality of the proposed system through evaluations and user studies.

The contributions of the proposed system are as follows.
\begin{itemize}
    \item We propose a novel visual analytics system that resolves DQ issues and enables optimal DQI procedures.
    \item We incorporate DQ evaluations with DQ dimension and K-S test into the visual analysis process to reduce reliance on expertise.
    \item We strategically integrate visual reasoning with statistical preprocessing to support more effective decision-making.
\end{itemize}
\section{Related Work}
\label{sec:related work}
DQ indicates whether data meets the requirements for its intended utilization~\cite{sebastian2012measuring}. However, since each user has unique requirements, the definition of DQ may vary depending on the users' objectives~\cite{batini2014opening}. Therefore, the DQ dimension has been proposed as a standardized metric for evaluating DQ~\cite{batini2006data}. The DQ dimension is a standardized criterion for evaluating user requirements, and an aggregated score of numerous DQ dimension estimates DQ. This section outlines the techniques used to improve DQ and the VA methods that support them. A comprehensive survey by Wang et al.~\cite{wang2024visual} systematically organizes and analyzes VA research for machine learning models from a data-centric perspective, covering data types, data-related tasks, and emerging research trends.

\subsection{Data Quality Improvement}
\label{subsec:mdqi}
Recognizing and addressing problems that negatively impact DQ is essential for enhancing DQ. DQI methods are categorized into: data- and process-driven strategies~\cite{batini2009methodologies}.

In ML, data preprocessing is a data-driven approach that uses imputation and feature selection techniques to improve DQ. Imputation is a technique that replaces missing values. It replaces missing values with a single value, such as mean, mode, or median, or with different values determined based on the correlation between instances, such as regression~\cite{zhang2016missing}. Feature selection filters out redundant or irrelevant features and are classified into filter and wrapper methods depending on when it is used in the model. The filter removes features below a certain threshold using correlation or mutual information. In contrast, the wrapper method selects a subset of features significantly impacting the model~\cite{chandrashekar2014survey}. Additionally, DQ are degraded by outliers, inconsistent values, and duplicate records. Outliers are identified and removed using statistical methods such as InterQuartile Range~(IQR), z-score, ML, and deep learning~\cite{hodge2004survey}. Inconsistent values and duplicate records are identified through a simple search and then deleted~\cite{garcia2016big}. These data preprocessing methods modify data uniformly without considering the intended usage by the users. Therefore, traditional data preprocessing is inappropriate for providing a comprehensive data-driven strategy to improve DQ.

The process-driven strategy is classified into process control and redesign. Process control is a technique that examines DQ to determine whether additional DQI is necessary. DQ is evaluated based on the rule method, uncertainty model, metrics method, and cost method in process control~\cite{bronselaer2021data}. The rule method measures the DQ established on specific rules, while the uncertainty model measures the DQ based on user input. The metrics method measures the DQ by calculating the similarity between instances, and the cost method measures the cost of DQI. On the other hand, process redesign is a technique for evaluating DQI procedures. In the DQI procedure, statistical tests such as the K-S, Shapiro Wilk, and Chi-Square tests are used as evaluation metrics~\cite{yeon2020visual, nakai2011simulation}. These statistical tests compare data before and after DQI. When the p-value is over than 0.05, it indicates that the DQI procedure does not distort the characteristics of the original data~\cite{arnold2011nonparametric}.

Each data has different characteristics, such as the number of DQ issues or their distributions, and the DQI method differs accordingly. For example, among the imputation techniques, the mean is fast and straightforward, but it may cause data bias, making it suitable only when there are few DQ issues. On the other hand, expectation maximization~(EM) accurately estimates statistical values of data but takes more time to calculate as the number of DQ issues increases~\cite{musil2002comparison}. Furthermore, the same DQI method can produce different performance enhancements depending on the ML model~\cite{li2019cleanml}. Therefore, it is crucial to identify suitable DQI methods and configure their orders to achieve optimal results. In this paper, we present a VA system that provides data-driven strategies essential for machine learning and assists users in identifying suitable DQI through process-driven strategies. The data- and process-driven strategies used in our system are introduced in Section~\ref{sec:data driven strategy} and Section~\ref{sec:process driven strategy}.

\begin{table*}[t]
    \small
    \caption{
        Summary of comparison between our system and previous studies.
    }
    \centering
    \label{Table:compRW}
    \begin{tabular}{p{0.05\textwidth}p{0.05\textwidth}p{0.2\textwidth}p{0.2\textwidth}p{0.2\textwidth}}
        \toprule
            \multicolumn{1}{c}{\textbf{System}} & \multicolumn{1}{c}{\textbf{Level}} & \multicolumn{1}{c}{\textbf{Target Issue}} & \multicolumn{1}{c}{\textbf{Visualization}} & \multicolumn{1}{c}{\textbf{User Interaction}}\\ 
        \midrule
            \multicolumn{1}{c}{\cite{xiang2019interactive}} & \multicolumn{1}{c}{Instance} & \multicolumn{1}{c}{Label error detection} & \multicolumn{1}{c}{t-SNE} & \multicolumn{1}{c}{Trusted item selection}\\
            \multicolumn{1}{c}{\cite{chen2024enhancing}} & \multicolumn{1}{c}{Instance} & \multicolumn{1}{c}{Temporal misalignment} & \multicolumn{1}{c}{Storyline} & \multicolumn{1}{c}{Alignment \& result correction}\\
            \multicolumn{1}{c}{\cite{yang2022diagnosing}} & \multicolumn{1}{c}{Instance} & \multicolumn{1}{c}{Low-quality shot and learner} & \multicolumn{1}{c}{Matrix and scatterplot view} & \multicolumn{1}{c}{Subset selection and feedback loop}\\
            \midrule
            \multicolumn{1}{c}{\cite{chen2021oodanalyzer}} & \multicolumn{1}{c}{Feature} & \multicolumn{1}{c}{Out-of-Distribution sample detection} & \multicolumn{1}{c}{Grid layout, kNN} & \multicolumn{1}{c}{Neighborhood-based comparision}\\
            \multicolumn{1}{c}{\cite{yang2020diagnosing}} & \multicolumn{1}{c}{Feature} & \multicolumn{1}{c}{Concept drift} & \multicolumn{1}{c}{Streaming scatterplot} & \multicolumn{1}{c}{Temporal trend interpretation}\\
            \multicolumn{1}{c}{\cite{gou2021vatld}} & \multicolumn{1}{c}{Feature} & \multicolumn{1}{c}{Accuracy and robustness} & \multicolumn{1}{c}{Semantic projection landscape} & \multicolumn{1}{c}{Semantic adversarial validation}\\
            \midrule
            \multicolumn{1}{c}{\cite{zhang2023labelvizier}} & \multicolumn{1}{c}{Label} & \multicolumn{1}{c}{Weak label auditing} & \multicolumn{1}{c}{Notebook-style UI} & \multicolumn{1}{c}{Text-based label inspection}\\
            \multicolumn{1}{c}{\cite{yang2024interactive}} & \multicolumn{1}{c}{Label} & \multicolumn{1}{c}{Label uncertainty} & \multicolumn{1}{c}{Bipartite graph} & \multicolumn{1}{c}{Importance reweighting}\\
            \multicolumn{1}{c}{\cite{chen2021interactive}} & \multicolumn{1}{c}{Label} & \multicolumn{1}{c}{GSSL graph refinement} & \multicolumn{1}{c}{Bybrid visualization} & \multicolumn{1}{c}{Manual graph structure adjustment}\\
            \multicolumn{1}{c}{\cite{sperrle2019viana}} & \multicolumn{1}{c}{Label} & \multicolumn{1}{c}{Subjective labeling} & \multicolumn{1}{c}{Layered Text/Graph} & \multicolumn{1}{c}{Annotation recommendation}\\
            \multicolumn{1}{c}{\cite{chen2022towards}} & \multicolumn{1}{c}{Label} & \multicolumn{1}{c}{Caption-based object labeling} & \multicolumn{1}{c}{Node-link, clustering} & \multicolumn{1}{c}{Label/object mutual refinement}\\
            \midrule
            \multicolumn{1}{c}{\multirow{2}{*}{\cite{hong2023how}}} & \multicolumn{1}{c}{\multirow{2}{*}{All}} & \multicolumn{1}{c}{\multirow{2}{*}{Missing, outliers, inconsistent, bias}} & \multicolumn{1}{c}{Battery chart, histogram, } & \multicolumn{1}{c}{Custom-built actions, }\\
             & & & \multicolumn{1}{c}{scatterplot, heatmap} & \multicolumn{1}{c}{DQI combination selection}\\
            \multicolumn{1}{c}{Ours} & \multicolumn{1}{c}{All} & \multicolumn{1}{c}{Multi-level DQ diagnosis and refinement} & \multicolumn{1}{c}{Explorative linked views} & \multicolumn{1}{c}{Feedback-based refinement}\\
        \bottomrule
    \end{tabular}
\end{table*}
\subsection{Visual Analytics for Data Quality Improvement}
Most VA systems for DQI provide automated analysis and interactive visualizations. In this section, we classify and review the VA systems for DQI into three levels~\cite{yuan2021survey}, including instance-, feature-, and label-levels.

Instance-level VA systems identify and correct DQ issues, such as missing values, outliers, and inconsistencies. These systems enable users to locate and correct instance-level errors directly, thereby enhancing the overall effectiveness of anomaly resolution. Kandel et al.~\cite{kandel2012profiler} propose a Profiler that allows users to analyze DQ issues at the instance level. The Profiler utilizes clustering and statistical-based techniques to detect the root causes of DQ issues and presents the results through various VA techniques such as histograms, area charts, and bar charts. Alemzadeh et al.~\cite{alemzadeh2020visual} proposed KVIVID, explicitly dealing with missing value treatment. KVIVID supports multiple imputation by providing diagnostics for missing values and visualizing the imputation results in a prediction matrix. DataDebugger~\cite{xiang2019interactive} leverages trusted item propagation and t-SNE-based navigation to identify and correct mislabeled items. Bauerle et al.~\cite{bauerle2020classifier} propose a pipeline-based system to examine suspicious samples by analyzing discrepancies between predicted and true labels. FSLDiagnostor~\cite{yang2022diagnosing} iteratively improves few-shot learning performance by identifying low-accuracy learners and samples, using matrix and scatter plot views. ActLocalizer~\cite{chen2024enhancing} supports error correction and propagation in temporal sequences through storyline visualizations.

At the feature-level, VA systems aim to minimize redundancy between features and maximize the relevance between instances. Krause et al.~\cite{krause2014infuse} proposed INFUSE to rank features with feature selection and extraction techniques, cross-validation folds, and classification models and visualize them with radial glyphs. Zhao et al.~\cite{zhao2019featureexplorer} proposed FeatureExplorer, which helps users interactively select and extract features from a high-dimensional feature space and modify the model based on feature importance and the performance of a regression model. OoDAnalyzer~\cite{chen2021oodanalyzer} combines ensemble detection with a grid-based layout to identify out-of-distribution (OoD) instances that deviate from the training distribution. DriftVis~\cite{yang2020diagnosing} focuses on temporal changes in feature distributions, utilizing streaming scatterplots and drift-degree plots to reveal the onset and nature of concept drift. VATLD~\cite{gou2021vatld} evaluates model robustness by analyzing feature-level accuracy in a traffic light detection model.

At the label-level, VA systems directly correct noisy labels in the data or identify groups of similar instances to label the data. Liu et al.~\cite{liu2018interactive} proposed LabelInspect support crowdsourced labeled data validation. LabelInspect infers the labels of instances based on the learning from the crowdsourced annotation model and helps to discover uncertain labels by expressing the similarity between instances through visualizations such as confusion matrices or scatter plots. VASSL was proposed by Khayat et al.~\cite{khayat2019vassl} to identify groups with anomalous behaviors by showcasing similarities between instances from different viewpoints. LabelVizier~\cite{zhang2023labelvizier} identifies text instances with noisy or weak labels through notebook-based interactions. Yang et al.~\cite{yang2024interactive} employ bipartite graphs to model reweighting interactions between training and validation samples, allowing for the dynamic reassignment of sample importance based on model performance. DataLinker~\cite{chen2021interactive} facilitates the interactive construction of high-quality graphs for semi-supervised learning by analyzing node-edge structures. VIANA~\cite{sperrle2019viana} supports argumentation annotation in subjective domains through progressive recommendations and layered visualizations, guiding users in the label-creation process. MutualDetector~\cite{chen2022towards} provides a pipeline for iteratively enhancing caption-based object label extraction and object detection.

Previous VA systems have primarily targeted specific data levels to address distinct aspects of DQ, employing interactions and visualizations suited to their respective scopes, as summarized in Table~\ref{Table:compRW}. In particular, the system proposed by Hong et al.~\cite{hong2023how} integrates both data- and process-oriented strategies to support multi-level DQ analysis through customized visual components and user-guided tasks. However, these approaches predominantly emphasize exploratory analysis and semi-automated DQI combination selection, lacking a generalizable feedback loop that tightly couples data cleaning with model performance enhancement. In contrast, our system unifies instance-, feature-, and label-level diagnostics with both data- and process-based strategies within an iterative, feedback-driven framework. This enables users to interpret quality issues using DQ dimensions and issue types comprehensively and to improve their data based on performance-guided feedback.
\section{Design Considerations}
\label{sec:design}
In this section, we outline the design of d-DQIVAR to provide optimal DQI procedures, addressing both current DQI challenges~\cite{alshdaifat2021effect, batini2009methodologies, li2019cleanml, musil2002comparison} and shortcomings of the most recent system~\cite{hong2023how}.

\subsection{Challenges of the DQI}
\label{subsec:cod}
Previous research has introduced DQI systems~\cite{yuan2021survey}. These systems typically focus on addressing specific DQ issues to enhance the performance of ML models~(\textbf{C1}). However, achieving comprehensive DQI is challenging as data encompasses instance-level, feature-level, and label-level DQ issues~\cite{alshdaifat2021effect}. Moreover, existing systems primarily focus on addressing quality issues through data preprocessing~(\textbf{C2}). However, since the appropriate DQI methods vary depending on the data or ML model, users encounter challenges in determining which DQI method to employ in specific circumstances~\cite{li2019cleanml}. Furthermore, certain studies opt to uniformly preprocess common DQ issues such as missing values or outliers~(\textbf{C3}). However, without granular adjustments, they may fail to produce data with the desired quality tailored to user requirements~\cite{batini2014opening}.

\subsection{Limitations of the existing system}
\label{subsec:los}
While the state-of-the-art (SOTA) system introduced by Hong et al.~\cite{hong2023how} embraced both data-driven and process-driven approaches to address prevailing challenges~(\textbf{C1}, \textbf{C2}, \textbf{C3}), it exhibited limitations. Specifically, the system suffered from diminished usability because of its demand for expert knowledge and the absence of visual reasoning. Their system was geared towards enhancing ML model performance through an extensive array of DQI methods. Therefore, the system has attempted to solve DQ issue by computing all possible cases of DQI methods and sequences on behalf of the users. While well-intentioned, this approach inadvertently led to user overwhelm. Consequently, the multitude of DQI choices can lead to user confusion~(\textbf{L1}). Also, some of the computed DQIs are likely to distort data characteristics. Choosing DQIs without causing data distortion proved challenging for users. This limitation restricted users only to following the unilateral analysis process suggested by the system~(\textbf{L2}). Moreover, the visual analysis process introduced in the existing study~\cite{hong2023how} risks overwhelming users by conveying information implicitly~(\textbf{L3}). The system complicates the task for users with limited expertise to discern valuable insights from the visualization. This limitation curtails the system's usability and prolongs the analysis duration.

\subsection{Design Goals}
This work articulates the design goals of d-DQIVAR, which aims to address the DQI challenges in Sections~\ref{subsec:cod} and limitations of the SOTA system in Section~\ref{subsec:los} by strategically integrating visual reasoning with statistical preprocessing.

\textbf{DG1: Comprehensive Scope of DQI.} Ensuring a comprehensive representation that encapsulates a wide range of DQ challenges addresses the narrowed perspective of existing DQI systems~(\textbf{C1}). We need to provide diverse DQI methods and their applicability to different data and ML models. We also need to reduce the challenge of decision-making in diverse situations~(\textbf{C2}). The system should guide users in choosing the appropriate DQI method tailored to the specific scenario.

\textbf{DG2: Expert Knowledge Integration \& Visual Reasoning.} We need to enhance the user experience by limiting the overwhelming number of DQI methods presented~(\textbf{C3}). By introducing a structured and intuitive interface, users should be able to navigate through DQI methods seamlessly. Also, the system needs to embed expert knowledge, ensuring even users without expertise benefit from best practices and expert insights~(\textbf{L1}). We need to enhance the visual reasoning capabilities of the system to reduce the time and effort traditionally required for optimal DQI procedure determination.

\textbf{DG3: Data Integrity \& User Autonomy.} We need to recognize and address the potential for data distortion inherent in some DQIs~(\textbf{L2}). The system should prioritize data integrity by guiding users away from DQIs that cause such distortions. Furthermore, the system should empower users with the autonomy to customize DQI processes while preserving data fidelity, thereby striking a balance between customization and DQ.

\textbf{DG4: Enhanced Usability Through Simplification.} The system needs to refactor the visual analysis process to provide users with explicit information~(\textbf{L3}). By decluttering visual presentations, the system aims to offer more user-friendly insights, thereby reducing analysis time and boosting usability.

Building on the development of the previous system~\cite{hong2023how}, the design goals of d-DQIVAR were iteratively refined and validated over an 11-month collaboration with six AutoML practitioners from diverse domains, including data processing, industrial analytics, healthcare analysis, and meteorological forecasting. These practitioners did not use standardized DQI tools; instead, they relied on ad hoc Python scripts, leading to limitations in consistency, scalability, and reproducibility. Through this collaboration, practitioners emphasized the need for scalability and adaptability across heterogeneous datasets and model types (\textbf{DG1}), intuitive visual guidance and expert-driven decision support (\textbf{DG2}), careful handling to avoid unintended data distortion and preserve domain-specific nuances (\textbf{DG3}), and simplified visual workflows to reduce cognitive load (\textbf{DG4}). These practitioner-driven insights ensured that the system aligns with real-world analytical needs and the practical constraints faced by domain experts.
\section{Data Quality Improvement Approach}
In this section, we delineate DQ and elucidate the data- and process-driven strategies employed by d-DQIVAR.

\subsection{Definition of Data Quality}
\label{sec:data quality definition}
We define DQ and DQ dimensions based on the research by Gupta et al.~\cite{gupta2021data}. Specifically, DQ is conceptualized as a metric assessing the appropriateness of data for ML applications. We also provide a comprehensive overview of DQ dimensions and associated issues. Table~\ref{Tab:definition} presents a summary of relevant information.

 \begin{table}[tb]
    \small
    \caption{
    The terminology used in the paper. The table presents definitions for DQ, DQ dimensions, and DQ issues. Also, it compares data- and process-driven strategies.
    }
    \centering
    \label{Tab:definition}
    \begin{tabular}{p{0.18\linewidth}p{0.18\linewidth}p{0.42\linewidth}}
    \toprule[.1em]
        \multicolumn{2}{c}{\textbf{Terminology}} & \multicolumn{1}{c}{\textbf{Definition}}\\
    \midrule[.1em]
        \multicolumn{2}{c}{Data quality} & \begin{tabular}[c]{@{}l@{}}Indicator whether data meets\\the requirements for its\\intended utilization\end{tabular} \\ \hline
        \multicolumn{2}{c}{Data quality dimension} & \begin{tabular}[c]{@{}l@{}}Standardized criterion for\\assessing data quality\end{tabular} \\ \hline
        \multicolumn{2}{c}{Data quality issue} & \begin{tabular}[c]{@{}l@{}}Problems that have a negative\\impact on data quality\end{tabular} \\ \hline
        \multirow{2}{2.5cm}{\begin{tabular}[c]{@{}l@{}}Data quality\\improvement\\method\end{tabular}}
        & \begin{tabular}[c]{@{}l@{}}Data-driven\\strategy\end{tabular} & \begin{tabular}[c]{@{}l@{}}Identifying and fixing\\ data quality issue\end{tabular} \\
        \cmidrule(lr){2-3}
        \multicolumn{1}{c}{} & \begin{tabular}[c]{@{}l@{}}Process-driven\\strategy\end{tabular} & \begin{tabular}[c]{@{}l@{}}Assessing data quality\\and designing data quality\\improvement procedure\end{tabular} \\
    \bottomrule[.1em]
    \end{tabular}
\end{table}

\subsubsection{Data Quality Dimension}
\label{subsubsec:dqd}
The DQ dimension serves as a standardized criterion for assessing DQ~\cite{batini2006data}. The DQ dimensions are rated on a scale of 0 to 100, with higher scores indicating better DQ. In this study, we map the DQ dimension to data completeness, outlier detection, data homogeneity, data duplicates, correlation detection, and feature relevance. These dimensions correspond to DQ issues such as missing values, outliers, inconsistent entries, duplicate records, high correlation, and low correlation, respectively. Since our system is designed explicitly for regression models, we do not incorporate class overlap, label purity, class parity, or data fairness into our DQ dimension.

Data completeness indicates whether enough data is available for the ML model to produce results accurately. Outlier detection indicates the data accuracy and users' trust in the data. Data homogeneity refers to how much data formats are consistent. The primary contributors to the deterioration of machine learning model performance are instances that are either missing or blank, instances exhibiting significant deviations from the norm, and instances with a distinct format compared to others, as highlighted in a study by Vinisha et al.~\cite{vinisha2022study}. We refer to these instances as missing values, outliers, and inconsistent values. Therefore, we calculate the ratio of missing values, outliers, and inconsistent values and represent them as data completeness, outlier detection, and data homogeneity, respectively. Data duplicates refer to how much data consists of different records. Duplicate records waste memory and computation time and result in imbalanced data. Therefore, we calculate the ratio of duplicate records and represent it as data duplicates. Correlation detection indicates how many features are independent of each other. Overfitting of the ML model can occur when the data contains features that are highly correlated with each other, which is known as multicollinearity~\cite{lieberman2014precise}. Therefore, we calculate the ratio of features that have a high correlation with each other in the data and represent it as correlation detection. Feature relevance refers to the extent to which the data consists of important features. Feature relevance is especially useful when working with high-dimensional data that has a large number of features in the ML model. We calculate the ratio of features that have a high correlation with a target feature in the data but have a low correlation with other features and represent it as feature relevance. Note that high correlation means features that have a high correlation with each other, and low correlation means features that have a high correlation with the target feature but a low correlation with other features.

\begin{figure}[tb]
	\centering
	\includegraphics[width=\linewidth]{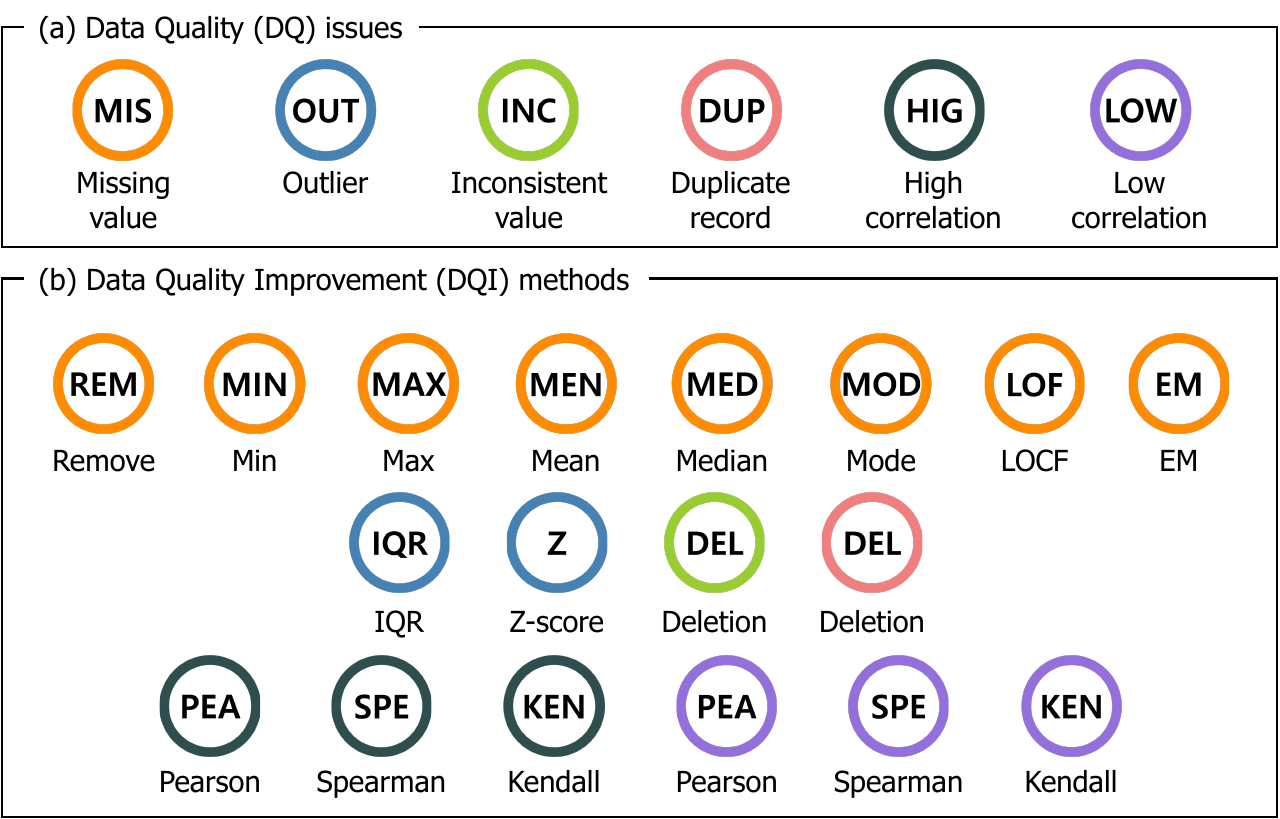}
	\caption{
        The icons illustrate the DQ issues in (a) and DQI methods in (b).
	}
	\label{fig:icon}
\end{figure}

\subsubsection{Data Quality Issues}
\label{subsubsec:dqi}
As referenced in Section~\ref{subsec:mdqi}, DQ issues detrimentally impact DQ. Our system detects DQ issues such as missing values, outliers, inconsistent entries, duplicate records, high correlation, and low correlation. A simple search easily detects missing values, inconsistent values, and duplicate records to identify DQ issues. However, it is necessary to establish criteria for detecting outliers and assessing correlations between features. In our approach, we employ the Interquartile Range (IQR) and z-score as criteria for identifying outliers, categorizing instances larger or smaller than 1.5 times the IQR and instances with a z-score greater than 3 as outliers, following the guidelines presented by Hodge and Austin~\cite{hodge2004survey}. Additionally, we utilize Pearson, Spearman, and Kendall correlation coefficients to quantify the relationships between features. A high correlation is defined when the absolute value of the correlation coefficient between features exceeds 0.8. This threshold follows widely accepted statistical guidelines, as correlation coefficients above 0.8 or below -0.8 typically indicate strong linear relationships and suggest potential redundancy among features~\cite{shrestha2020detecting}. Conversely, correlations with absolute values between -0.2 and 0.2 are considered low, reflecting negligible or weak linear associations~\cite{hinkle2003applied}. These thresholds provide a practical basis for feature analysis, facilitating the identification and management of redundant or weakly related variables. Furthermore, recognizing that appropriate threshold values may differ by domain, d-DQIVAR provides a configurable threshold editing feature to support domain-specific adjustments by experts. We represent DQ issues using icons as illustrated in Figure~\ref{fig:icon}~(a).

\begin{figure*}[tb]
	\centering
	\includegraphics[width=\linewidth]{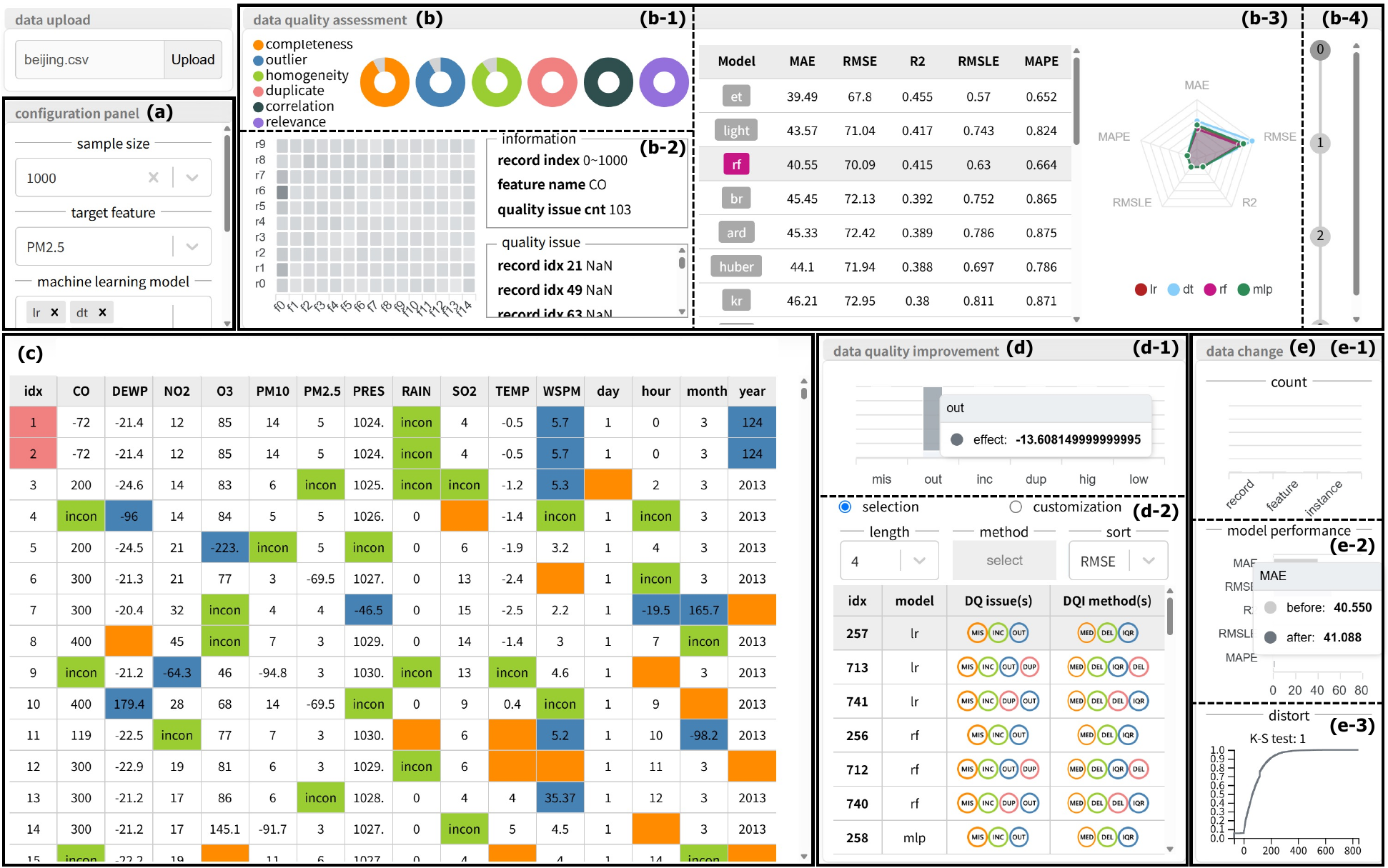}
	\caption{
	Our visual analytics system. (a) is a configuration panel for selecting parameters. (b) is a DQ assessment view that analyzes the DQ for each DQI procedure step. (c) is a data overview showing DQ issues in detail on an instance, record, and feature levels. (d) is a DQI view in which the DQ is improved by selecting DQI procedures or customizing specific parts of them. (e) is a data change view summarizing the difference before and after DQI.
	}
	\label{fig:system}
\end{figure*}

\subsection{Data-driven Strategy}
\label{sec:data driven strategy}
To resolve DQ issues, we apply modification methods, such as imputation, deletion, and feature selection. Imputation ensures data completeness by addressing missing values. Remove, min, max, mean, median, mode, last observation carried forward~(LOCF), and EM are used as imputation methods. We also remove outliers, inconsistent values, and duplicate records to ensure outlier detection, data homogeneity, and data duplicates. Feature selection ensures correlation detection and feature relevance by removing features having a high correlation with each other or a high correlation with target features but a low correlation with other features. The system resolves the narrow perspective of existing systems by comprehensively addressing DQ enhancement methods (\textbf{DG1}).

\subsection{Process-driven Strategy}
\label{sec:process driven strategy}
Enhancing DQ involves the iterative implementation of data-driven strategies. Consequently, it becomes essential to complement this approach with a process-driven strategy that evaluates and formulates DQ procedures.

Our system assesses the DQ using the DQ dimension and the performance of the ML models. The DQ dimension serves as a measure of how many DQ issues are present in the data. The performance is used to gauge the data suitability for ML models. We use the autoML library pycaret~\cite{PyCaret} to calculate the ML model performance through cross-validation.

The system generates DQI procedures to suggest optimal DQI to the users. The DQI procedures include all cases configured with different DQI methods and sequences. As we deal with six types of DQ issues, the DQI procedure consists of at least one and at most six DQI methods. The system then generates quality-improved data for each DQI procedure. Several procedures of DQI methods may potentially enhance DQ but also introduce distortions in data characteristics. To mitigate this, we exclusively consider DQI procedures that exhibit a p-value of over than 0.05 in the K-S test for both the original data and the data after DQI. Unlike other statistical tests, the K-S test does not make assumptions about a specific data distribution. This flexibility allows us to apply the K-S test to a diverse range of datasets. Following this, the system computes the performance of the machine learning models for each dataset after quality improvement. To assist users in their decision-making process, the system recommends the DQI procedure demonstrating the most significant improvement in the ML model performance~(\textbf{DG1}).
\section{VA System for Data Quality Improvement}
\label{sec:system}
d-DQIVAR integrates multiple coordinated views into a unified analytical workflow. Heatmaps facilitate the early detection of problematic data regions, radar charts support comparative evaluation of model performance, and matrix views provide detailed representations of feature correlations and DQ metrics. This coherent coupling enhances analytical efficiency while reducing cognitive overhead for users. From this foundation, the following section describes how d-DQIVAR improves DQ through a cohesive set of visual and analytical strategies. Additionally, we describe the computing server responsible for processing information within the system and provide an overview of the system views as presented in Figure~\ref{fig:system}. 

\begin{figure*}[tb]
	\centering
	\includegraphics[width=\linewidth]{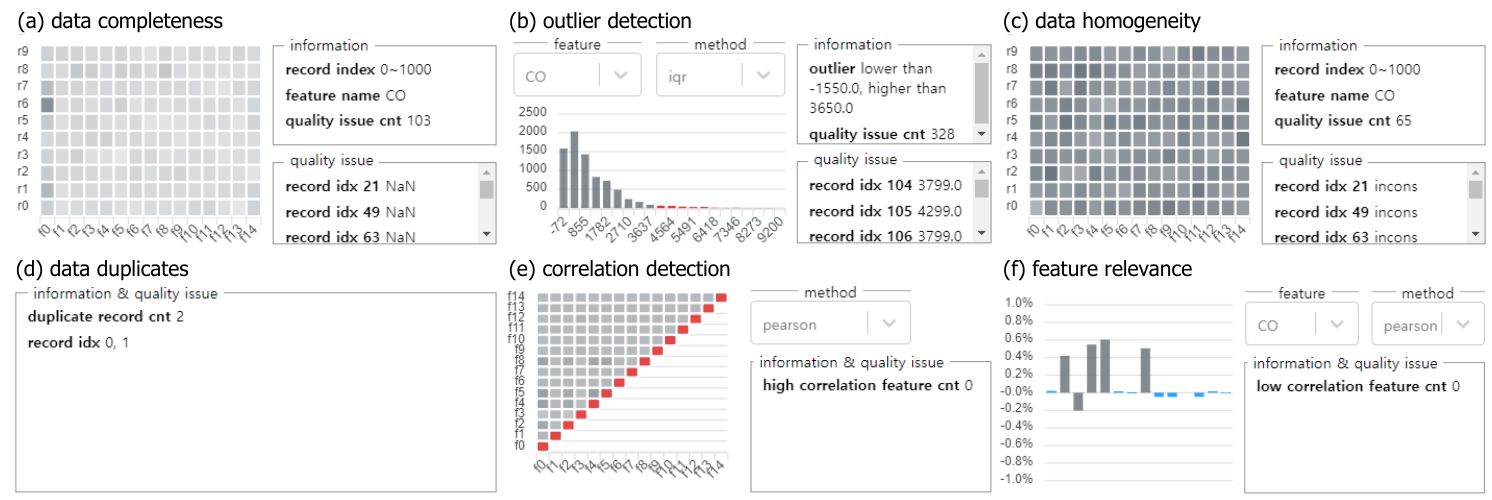}
	\caption{
        d-DQIVAR provides different VA techniques and information according to the DQ dimension selected in the donut chart.
	}
	\label{fig:reasoning}
\end{figure*}
\subsection{Data Quality Improvement Process}
\label{subsec:DQIP}
The system has been developed using React.js, D3.js, and Apexchart.js. The system takes data and parameters from the data upload and configuration panel and forwards them to the computing server. The computing server has been developed using Python-flask. As explained in Section~\ref{sec:process driven strategy}, the computing server calculates the DQI procedures, the quality-improved datasets, and the performance of the ML models. The computed results are then transferred back to the system, which facilitates the user's decision-making by visualizing the essential information to evaluate and improve the DQ.

The system offers various DQI methods and ML models. Therefore, much cost is spent on computing DQI procedures, quality-improved datasets, and the performance of the ML models. To address this problem, we devised a system so that the server calculates and sends them back to the system.

The cost depends on the number of DQ issue types, ML models, and the data size. We default set using data consisting of 15 features, 10,000 records, and six types of DQ issues with 5 ML models. Afterward, we changed the number of DQ issue types, number of ML models, and data size from the default settings. When the number of DQ issue types increases from 1 to 6, the required time increases from about 5 minutes to about 27 hours. When the number of ML models increases from 1 to 25, the required time increases from about 24 hours to about three days. When the data size increases from 500 records to 10,000 records, the required time increases from about 4.5 hours to about 26 hours. The time complexity is $O(n)$ for the number of records, $O(n^2)$ for the number of features, and $O(n!)$ for the number of DQ issue types.

\subsection{Selection of Data \& Machine Learning Models}
Figure~\ref{fig:system} (a) is the configuration panel that lets users specify target features, ML models, and evaluation metrics for DQI. d-DQIVAR provides a suite of 25 ML models, including LR, LASSO, RIDGE, EN, LAR, LLAR, OMP, BR, ARD, PAR, RANSAC, TR, HUBER, KR, SVM, KNN, DT, RF, ET, ADA, GBR, MLP, XGBOOST, LIGHTGBM, and LIGHTGBM. Additionally, it incorporates five evaluation metrics, including MAE, RMSE, R2, RMSLE, and MAPE.

\subsection{Data Quality Evaluation}
\label{sec:data quality assessment view}
Figure~\ref{fig:system}~(b) is a DQ assessment view that evaluates the DQ based on the DQ dimension and the performance of ML models. This view aims to reduce reliance on expert knowledge in assessing DQ and model performance by providing users with information about DQ dimension and machine learning model performance. (b-1) visualizes six DQ dimensions—completeness, outlier, homogeneity, duplicate, correlation, and relevance—using color-coded donut charts. The size of each segment reflects the corresponding dimension’s score, which is computed at the feature level and aggregated to the dataset level to support high-level overviews. Donut charts were selected for their effectiveness in conveying part-to-whole relationships, allowing users to quickly identify which dimensions contribute most to overall DQ degradation. This design choice minimizes the need for domain expertise during the initial assessment phase (\textbf{DG4}). (b-2) provides the VA technique and information to infer the DQ dimension. (b-3) displays the performance of each ML model using both a table and a radar chart. The table summarizes the results of 25 ML models across five evaluation metrics. Users can select up to five models, whose performance is then visualized in the radar chart. The radar chart overlays the selected models, with each axis representing a specific evaluation metric, enabling users to visually compare performance and identify trade-offs across models. The area of the radar chart is inversely proportional to the performance of the ML model, and the closer it is to the pentagon, the more uniform the performance distribution. The DQ assessment view tables and radar charts facilitate quick and easy comparison of ML model performance. This dual view facilitates the assessment of multi-objective models, enabling users to make informed decisions aligned with their task-specific priorities (\textbf{DG2}). (b-4) presents a DQI procedure tree that enables users to trace and analyze the incremental impact of each preprocessing step. This design supports process-aware evaluation, allowing users to isolate the effects of specific transformations and explore alternative DQI strategies. The tree abstraction thereby promotes transparency, reversibility, and iterative refinement of DQ interventions (\textbf{DG3}). The selected step in the tree diagram is highlighted in dark gray, and the system updates the views to analyze the data applied up to that step. Users analyze the DQ at each step of the DQI procedure by referring to intuitive charts and detailed explanations~(\textbf{DG4}).

\subsection{Visual Reasoning with Interactions}
To provide detailed information on the selected DQ dimensions in Figure~\ref{fig:system}~(b-1), the system expands Figure~\ref{fig:system}~(b-2) as presented in Figure~\ref{fig:reasoning}. Using these visual analysis methods and information, users can infer the causes of the DQ dimension score and consider ways to improve it~(\textbf{DG2}).

Figure~\ref{fig:reasoning}~(a) and (c) show data completeness and data homogeneity using heatmaps, respectively. We employ heatmaps to rapidly identify densely clustered areas of DQ issues from large volumes of data. The heatmap shows the distribution of DQ issues related to the DQ dimension, with the x-axis and y-axis representing features and records, respectively. To save space, 20 records were aggregated into each group, with each y-axis label (e.g., r0 to r9) representing a distinct group; thus, the heatmap summarizes a total of 200 records. The intensity of cell colors in the heatmap directly corresponds to the frequency of DQ issues, effectively directing users’ attention to problematic records and features. By simplifying complex data patterns, this heatmap reduces cognitive load and facilitates rapid decision-making within the system’s DQI workflow (\textbf{DG2}, \textbf{DG4}). Figure~\ref{fig:reasoning} (b) describes outlier detection using a histogram. We utilize histograms to delineate outlier criteria for each feature distinctly. The histogram shows the distribution of selected features, and the x-axis and y-axis represent the feature scale and the number of instances included in the scale, respectively. The x-axis of the histogram is aggregated into 20 sections. The histogram highlights outliers in red, and the height of the red bar is proportional to the number of outliers. Figure~\ref{fig:reasoning} (d) explains data duplicates by indicating the number of duplicate records and their index. The count of duplicate records is referenced numerically to minimize confusion without separate visualization. Figure~\ref{fig:reasoning} (e) employs a correlation matrix to visualize pairwise relationships between features and facilitate their comparison. In the correlation matrix, the axes represent the features. Feature pairs exhibiting high correlation are highlighted in red, with their count indicating the prevalence of strong linear relationships within the data. Figure~\ref{fig:reasoning} (f) shows feature relevance using a positive-negative bar chart. We employ a positive-negative bar chart to facilitate an easy understanding of the important features by representing both the magnitude and sign of correlation coefficients. The positive-negative bar chart shows the correlation between other features based on the selected feature, and the x-axis and y-axis represent the correlations with the features. In the positive-negative bar chart, a high correlation is highlighted in red, and a low correlation is highlighted in blue. If the bar corresponding to the target feature is red and the other bars are blue, the selected feature causes a feature relevance problem.

\subsection{Checking Actual Data with DQ Issues}
\label{sec:data overview}
Figure~\ref{fig:system}~(c) provides a comprehensive data overview by visualizing the distribution of DQ issues across features and instances. Each cell encodes the presence and intensity of a specific DQ issue using the color scheme introduced in Figure~\ref{fig:system} (b-1), ensuring perceptual consistency across views. This view dynamically updates based on the data state selected in Figure~\ref{fig:system} (b-4), allowing users to examine the localized effects of DQI procedures. By aggregating and rendering DQ issues at the instance-feature level, the visualization supports the discovery of patterns such as feature-specific anomalies, row-level data corruption, and co-occurring issues across multiple dimensions. This design facilitates fine-grained diagnosis of problematic regions and guides targeted refinement efforts (\textbf{DG4}).

\begin{figure*}[tb]
	\centering
	\includegraphics[width=\linewidth]{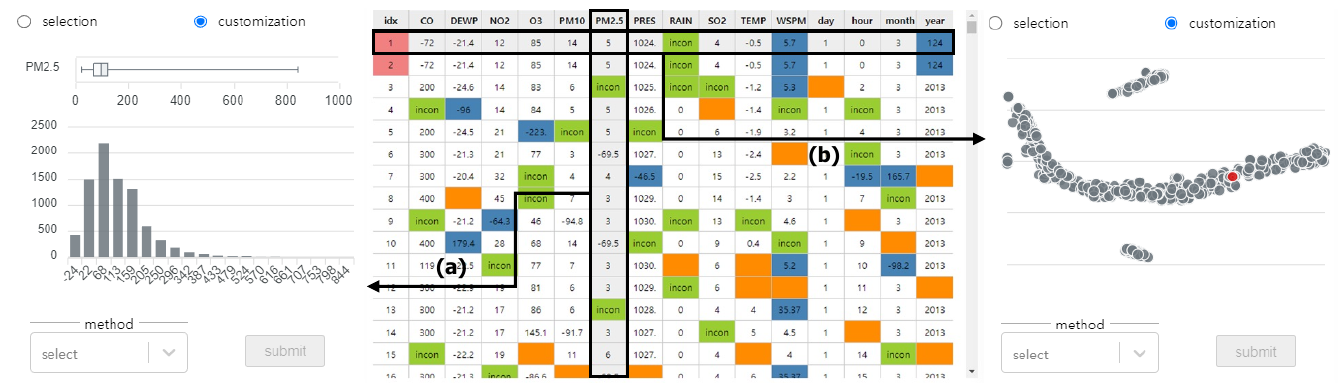}
	\caption{
        Views for customizing DQI. d-DQIVAR provides different VA techniques and information depending on the feature and record selected in the data overview. (a) presents feature-level distribution analysis through box plots and histograms, while (b) provides instance-level inspection using a t-SNE-based scatter plot.
	}
	\label{fig:improvement}
\end{figure*}

\subsection{DQI Procedure Selection}
\label{sec:data quality improvement view}
Figure~\ref{fig:system}~(d) is a DQI view. This view provides information required to improve DQ. (d-1) quantifies the impact of individual DQ issues on ML model performance by comparing results before and after issue-specific DQI applications. Each bar represents the average performance difference across models and evaluation metrics, grouped by DQ dimension. This visualization highlights the data issues that most significantly affect predictive accuracy and the corrections that yield the most significant improvement, thereby guiding users toward more effective and impactful DQI strategies (\textbf{DG2}). (d-2) presents DQI procedures derived from a systematic evaluation of multiple preprocessing sequences, ranked according to user-defined evaluation metrics to support both performance-driven and constraint-aware decision-making. The system enables users to adjust constraints on method types and sequence lengths, providing flexibility in generating pipelines that range from conservative to comprehensive strategies (\textbf{DG1}). Fine-grained control over procedure structure promotes transparency and helps balance data fidelity with transformation effectiveness, reducing the risk of unintended semantic distortion (\textbf{DG3}). Metric-specific ranking logic further guides users in selecting DQI pipelines aligned with their performance objectives (\textbf{DG2}).

\subsection{Customizing DQI Procedures with Interactions}
The system provides interactive features such as Figure~\ref{fig:improvement} that allow users to customize the DQI procedures. The customization radio button in Figure~\ref{fig:system}~(d-2) activates this view. DQI procedures in d-DQIVAR are designed to foreground the user's role in selecting and refining DQ interventions. Rather than adopting a fixed, fully automated pipeline, the system enables iterative strategy refinement by integrating domain knowledge with visual cues from multiple coordinated views. This human-in-the-loop approach supports, rather than replaces, expert judgment, fostering informed and context-sensitive decision-making.

In this workflow, each view supports a distinct phase of the user's decision-making process. The data overview in Figure~\ref{fig:system}~(c) enables users to identify localized DQ issues by mapping them across features and records. Upon selecting a target feature or instance, the system generates diagnostic visualizations tailored to the selection. Box plots and histograms in Figure~\ref{fig:improvement}~(a) summarize statistical properties and reveal distributional anomalies, aiding decisions on interventions such as normalization or outlier removal. Histograms also expose skewness and modality, informing transformation or binning strategies. For instance-level analysis, the t-SNE scatter plot in Figure~\ref{fig:improvement}~(b) contextualizes the selected record within the global data structure, facilitating the detection of anomalous or cluster-inconsistent entries. Users can assess whether a record lies in a dense or sparse region, thereby evaluating its relative importance or redundancy.

Guided by diagnostic insights, users can apply, reorder, or omit DQI methods via the configuration panel. The radar chart in Figure~\ref{fig:system}~(b-3) then provides immediate visual feedback on the impact of each customized DQI sequence on ML performance. This coordinated workflow supports goal-oriented interaction, linking problem identification, hypothesis generation, method selection, and outcome evaluation within a unified reasoning loop.

\subsection{Comparison of Changes by DQI Procedure}
Figure~\ref{fig:system} (e) is a data change view that summarizes differences resulting from DQI using a bar chart, horizontal bar chart, and eCDF. (e-1) uses a bar chart to display changes in data size, including the number of features, records, and instances, resulting from DQI procedures. (e-2) is a horizontal bar chart that compares the ML model performance with the original data and quality-improved data in light gray and dark gray, respectively. The length in the horizontal bar chart is inversely proportional to the ML model performance. (e-3) shows the eCDFs of the original data and the quality-improved data in light gray and dark gray, respectively. The x-axis of eCDF denotes the feature scale, while the y-axis indicates the number of instances within the feature corresponding to the scale on the x-axis. Deviations between the two curves indicate potential distortions introduced by DQI, such as shifts in feature scale or instance-wise redistribution. By revealing unintended structural changes, this view allows users to verify whether the applied DQI procedures preserve the original data semantics, thereby reinforcing trust in the preprocessing pipeline. Consequently, these visual components reduce reliance on expert intuition by making critical data transformations and their side effects explicitly observable. This design empowers users to assess whether their DQI procedures introduce meaningful improvements while preserving essential data characteristics~(\textbf{DG3}).

\begin{figure}[t]
	\centering
	\includegraphics[width=\linewidth]{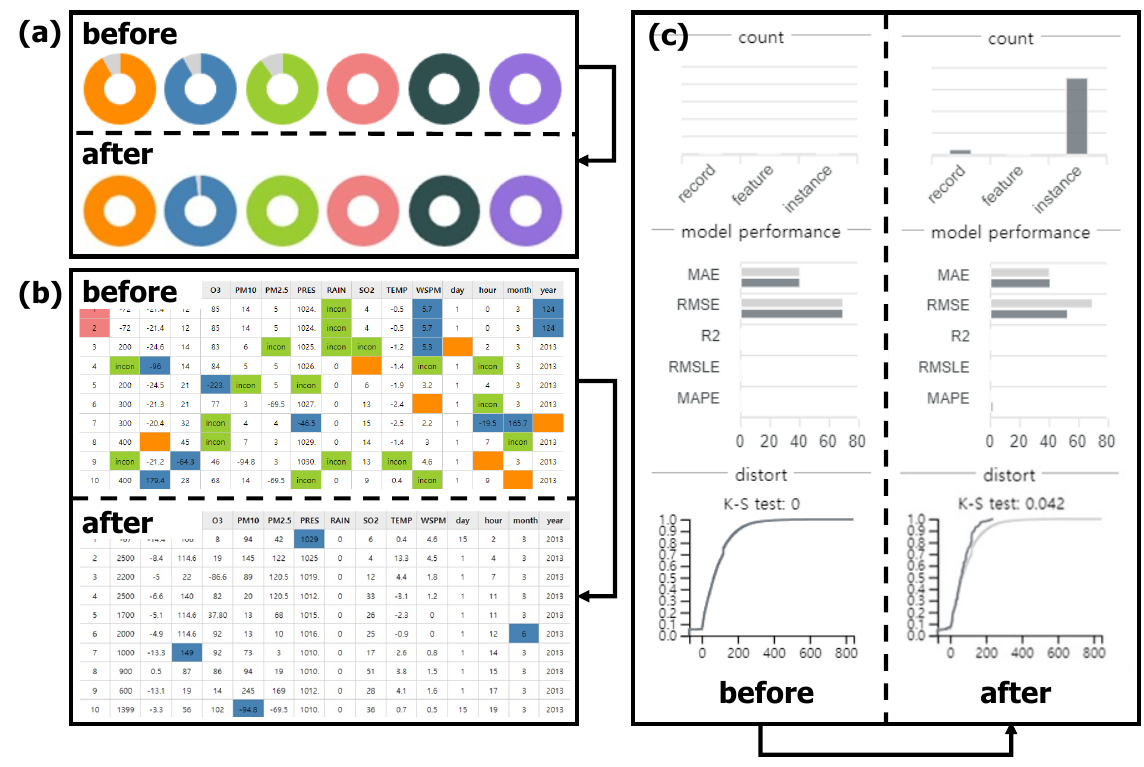}
	\caption{
		Comparison of before and after performing DQI procedures. (a) reveals that the DQ dimension scores are improved. (b) shows that DQ issues are reduced. (c) implies data size change, ML model performance change, and distortion degree of data characteristics.
	}
	\label{fig:after}
\end{figure}
\begin{figure*}[t]
	\centering
	\includegraphics[width=\linewidth]{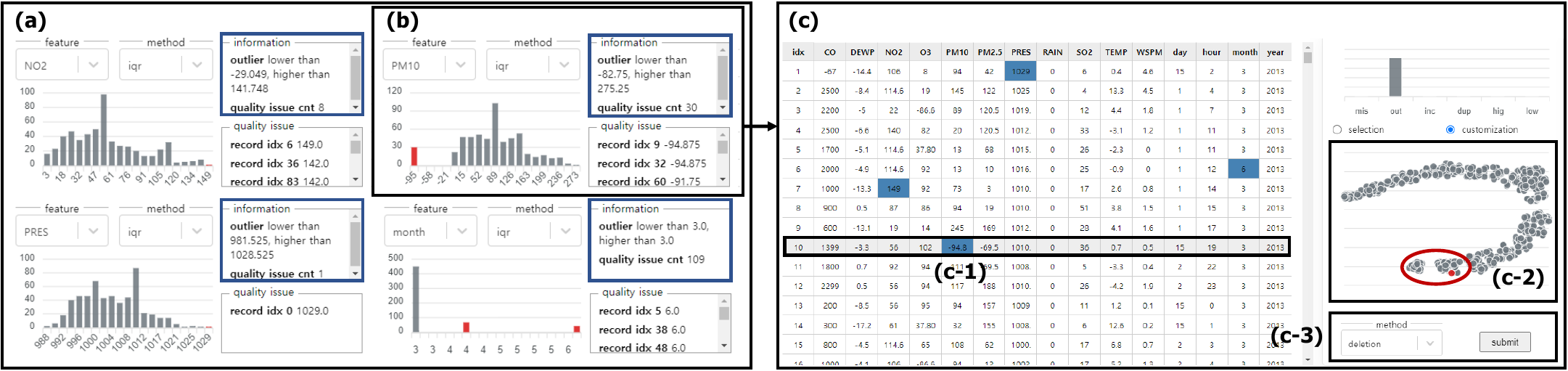}
	\caption{
		Customized DQI procedure. (a) shows features that still have outliers after DQI. (b) tells that the scale of \textit{PM10} is anomalous. (c) provides a detailed DQI by removing records with negative \textit{PM10} values.
	}
	\label{fig:custom}
\end{figure*}

\section{Case Study}
\label{sec:casestudy}
In this section, we present case studies with the system. We use the Beijing air quality dataset from the UCI ML repository and the Boston house price dataset from the scikit-learn library. The Beijing air quality dataset consists of 15 features and 10,000 records. We randomly generate a total of 30\% with missing values, outliers, and inconsistent values, duplicate the first and second records in the Beijing air quality dataset, and use them in Section~\ref{subsec:beijing}. Although the Beijing air quality dataset contains native DQ issues, we generated additional DQ issues to demonstrate the system effectively. The Boston housing price dataset consists of 14 features and 506 records. Unlike the Beijing air quality dataset, we use the origin Boston house price dataset in Section~\ref{subsec:house} without data poisoning.

\begin{figure*}[t]
	\centering
	\includegraphics[width=\linewidth]{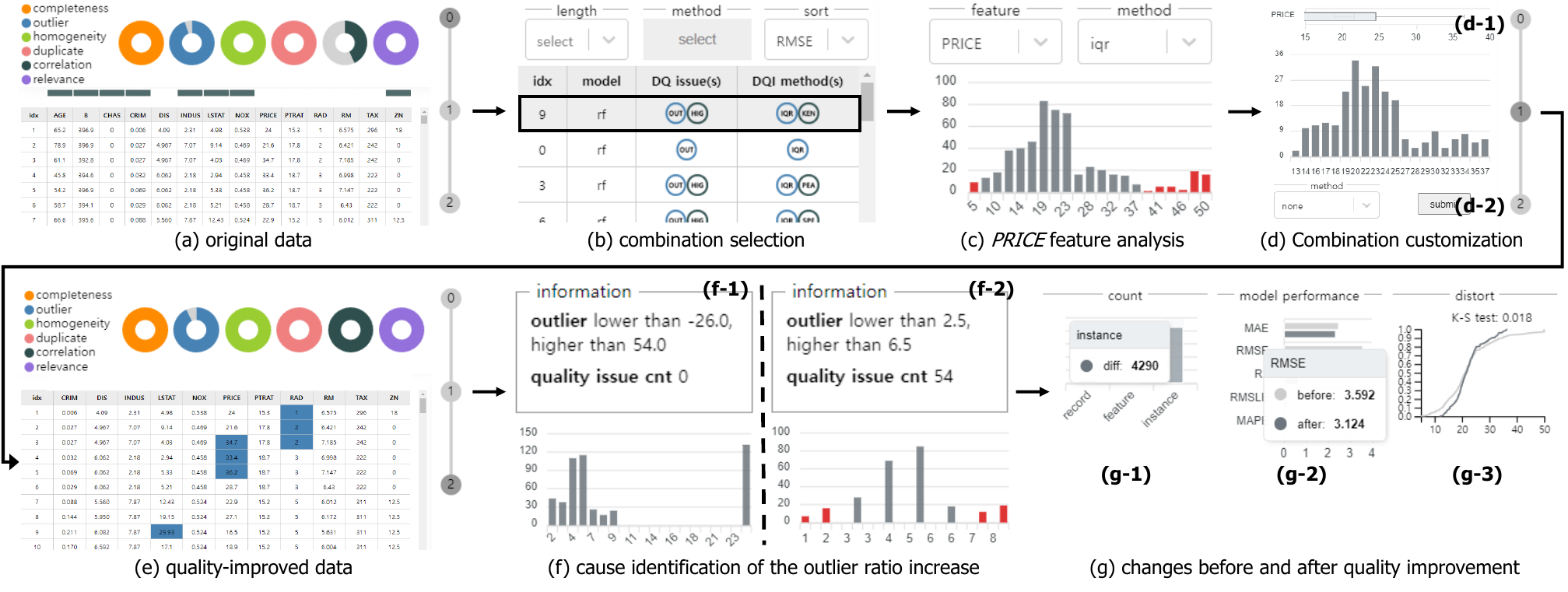}
	\caption{
		Regression for Boston house price. (a) is the original data. (b) displays the index 9 procedure with the lowest RMSE among all DQI procedures. (c) shows that \textit{PRICE} has a normal scale, but the system indicates that outliers exist. (d) shows that outliers present in \textit{PRICE} are not handled. (e) is the quality-improved data. (f) compares the data before and after DQI to show the cause of the outlier ratio increase. (g) summarizes the changes in the number of instances, RMSE, and p-value of the K-S test after the DQI.
	}
	\label{fig:house}
\end{figure*}

\subsection{Beijing Air Quality Dataset}
\label{subsec:beijing}
We aim to improve the quality of the Beijing air quality dataset for an ML model that regresses the concentration of fine particulate matter. We select \textit{PM2.5}, which means the concentration of fine particulate matter, as the target feature. We select LR, DT, RF, and MLP for ML models and RMSE and R2 for evaluation metrics~\cite{lin2021application}. We analyze the quality of the original data, select DQI procedures that enhance the performance of the ML model, check the DQ for each DQI procedure step, and improve the DQ in the order of customizing specific parts of the DQI procedures.

Figure~\ref{fig:system}~(b) represents the quality of the original data based on the DQ dimension and the performance of the ML model. In (b-1), the scores of the DQ dimension are visualized in the donut charts. Data completeness, outlier detection, and data homogeneity exhibit relatively low scores compared to other DQ dimensions. Therefore, we conduct a detailed analysis of data completeness, outlier detection, and data homogeneity as presented in Figure~\ref{fig:reasoning}~(a), (b), and (c). Figure~\ref{fig:reasoning}~(a) and (c) show that missing values and inconsistent values are distributed randomly. Figure~\ref{fig:reasoning}~(b) tells that outliers are consistent across all features. Therefore, we employ DQI methods to address data completeness, outlier detection, and data homogeneity. The performance of each ML model is visualized with the table and radar chart in Figure~\ref{fig:system}~(b-3). LR, RF, and MLP exhibit a similar performance distribution, but DT has higher MAE and RMSE than other ML models. Therefore, we remove DT from the analysis.

Figure~\ref{fig:system}~(c) shows the distribution of DQ issues within the original data in detail as a table. As confirmed in (b), many missing values, outliers, and inconsistent values are found in the data. In addition, there are no features that have a high correlation with each other or a high correlation with the target feature but a low correlation with other features. Additionally, we find that the index 1 and the index 2 records are duplicated.

Figure~\ref{fig:system}~(d-1) shows the effect on the ML model performance for each DQI method in a bar chart. Applying a DQI method related to the outlier to the original data resulted in a performance enhancement by 13.608. However, the performance difference before and after applying other DQI methods is negligible. Therefore, we deal with a DQI method related to outliers. We filter the DQI procedures based on the information acquired in Figure~\ref{fig:system}~(b), (c), and (d-1). Index 257, located at the top of the table after filtering, is the optimal DQI procedure. The index 257 procedure uses the LR model and solves DQ issues in the order of missing value, inconsistent value, and outlier. In addition, the index 257 procedure improves the DQ in the order of median, deletion, and IQR. We select 3, the last step of the index 257 procedure, in (b-4), to analyze the data after performing the DQI.

Figure~\ref{fig:after} compares the data before and after performing the index 257 procedure. (a) shows that data completeness, outlier detection, and data homogeneity scores are improved. (b) shows that missing values, inconsistent values, and duplicate records are removed, and only a small number of outliers exist. Note that the system re-evaluates the DQ based on the generated quality-improved data. Therefore, in the quality-improved data, outliers in the original data have been removed, but outliers identified as new criteria still exist. (c) shows the difference between the original data and the quality-improved data using a bar chart, horizontal bar chart, and eCDF. Due to the DQI, the number of records and instances has been reduced. The ML model performance was improved with a decrease in RMSE. The data characteristics were not distorted because the $p$-value of the K-S test was 0.042, and the two eCDF forms were similar to each other.

After applying the DQI procedure, the data still contains outliers that are identified by the new criterion. Therefore, we examine these outliers, as illustrated in Figure~\ref{fig:custom}. Outliers are observed in \textit{NO2}, \textit{PM10}, \textit{PRES}, and \textit{month} variables, which correspond to the mean concentration of NO2, concentration of PM10, barometric pressure, and the month. Figure~\ref{fig:custom}~(a) presents the per-feature distribution, outlier criterion, and the number of outliers. The system sets the outlier criterion per feature as the instances that fall below -29.049 or exceed 141.748 for \textit{NO2}, instances that fall below -82.75 or exceed 275.25 for \textit{PM10}, instances that fall below 981.525 or exceed 1028.525 for \textit{PRES}, and instances that are not equal to 3 for \textit{month}, as indicated by the blue highlighted area. Most of the outlier criteria specified by the system are possible scales. However, as shown in Figure~\ref{fig:custom}~(b), \textit{PM10} represents a concentration and cannot be negative. However, the histogram x-axis shows a range from -95 to 273. Therefore, we analyze records where \textit{PM10} is negative, as shown in Figure~\ref{fig:custom}~(c). We select the records where \textit{PM10} is negative as in (c-1) and locate them in (c-2). Records with negative \textit{PM10} are distributed mainly in the red highlighted area in the scatter plot. Therefore, we judge these as outliers and delete them, as shown in (c-3). Finally, we finish improving the DQ.

\subsection{Boston House Price Dataset}
\label{subsec:house}
We aim to improve the quality of the Boston house price dataset for a regression model. In the configuration panel, we set \textit{PRICE} as the target feature, RF, DT, and LR as the ML models, and RMSE as the evaluation metric~\cite{rawool2021house}. The DQI procedure proceeds in the order shown in Figure~\ref{fig:house}.

Figure ~\ref{fig:house}~(a) shows the DQ issues present in the original data using donut charts and tables. A small number of outliers are randomly distributed in the data. \textit{AGE}, \textit{B}, \textit{CHAS}, \textit{CRIM}, \textit{INDUS}, \textit{LSTAT}, \textit{NOX}, and \textit{ZN} have a high correlation with each other. Figure~\ref{fig:house}~(b) shows all DQI procedure calculated by the system. The system computes all cases of performing DQI methods that resolve issues for outliers and features that have a high correlation with each other. Therefore, DQI procedures consist of at most two steps. Based on RMSE, the optimal DQI procedure is the Index 9 procedure with the RF model in the order of outliers and features that have a high correlation with each other. In addition, the index 9 procedure uses IQR and Kendall as DQI methods.

House prices are subject to unusual events such as interest rates and rumors. Therefore, performing a batch of DQI methods for all outliers is risky. Figure~\ref{fig:house}~(c) shows the original \textit{PRICE} before resolving outliers in the index 9 procedure. The system detects outliers in the \textit{PRICE}, even though it has a normal scale ranging from 5 to 50. Therefore, we customize the first step of the index 9 procedure not to handle outliers in the \textit{PRICE}. Figure~\ref{fig:house}~(d) shows \textit{PRICE} after solving outliers in the index 9 procedure. As shown in the x-axis of (d-1), the \textit{PRICE} scale was changed from 13 to 37. Therefore, we select none in (d-2), meaning no DQI method is applied.

Figure~\ref{fig:house}~(e) shows the DQ issues that exist in the quality-improved data. After DQI, since \textit{AGE}, \textit{B}, and \textit{CHAS} are removed from the data, all features with high correlation with each other are solved. However, after DQI, the outlier ratio increased compared to the original data. Therefore, we analyze the cause of the increased outlier ratio. Figure~\ref{fig:house}~(f) shows the distribution, outlier criterion, and the number of outliers of \textit{RAD}, which means the index of accessibility to radial highways. Before the DQI, as shown in (f-1), the system detected instances with values smaller than -26 or greater than 54 as outliers. However, after the DQI, the system identifies instances with values less than 2.5 or greater than 6.5 as outliers, as illustrated in (f-2). Therefore, the number of outliers suggested by the system increased from 0 to 54. After DQI, since \textit{RAD} has a scale from 1 to 8, as seen on the histogram's x-axis, all the outliers identified by the system are normal. Therefore, we keep the outliers unprocessed.

Figure~\ref{fig:house}~(g) shows data characteristics changes after DQI. (g-1) shows that the number of instances decreases by 4290. (g-2) shows that the RMSE decreased from 3.592 to 3.124, and the ML model performance also improved for other evaluation metrics. (g-3) shows that based on the similarity of the two eCDF and a p-value, 0.018, from the K-S test, we can infer that the DQI procedure is appropriate even though a customized DQI procedure of DQI methods was used.
\section{Evaluation}
\label{sec:evaluation}
We evaluate the DQI procedures divided into DQ issue ratios and DQ issue types. We use the DQ dimension, the performance of the ML model, and the K-S test as evaluation indicators. Note that although the system offers 25 ML models and 5 evaluation metrics, in the evaluation, we use the LR, DT, and RF with RMSE commonly mentioned in Section~\ref{sec:casestudy} to compute the ML model performances with the bike sharing dataset and forest fires dataset provided by UCI ML repository, and the diabetes dataset provided by scikit-learn library. The bike sharing dataset consists of 14 features and 731 records, and the target feature is \textit{cnt}, the number of rental bikes. The Forest fires dataset consists of 11 features and 517 records, and the target feature is \textit{area}, the forest fire area. The Diabetes dataset consists of 11 features and 442 records, and the target feature is \textit{diabetes}, diabetes levels.

We poison the data with different DQ issues ratios and types for the evaluation. DQ issues at the instance level are included at the record and feature level. Therefore, we exclude duplicate records, features with a high correlation with each other, and features with a high correlation with the target feature but a low correlation with other features from the poisoning target. We create a total of 5 poisoning data per data to compare DQI procedures according to the DQ issue ratios. We increase the DQ issue ratio by 5\% from 10\% to 30\%~\cite{thomas2021systematic}. We treat only missing values, outliers, and inconsistent values as DQ issues to set the other factors equal except for the DQ issue ratio. In addition, we create a total of 7 poisoning data for each data to compare DQI procedures according to DQ issue types that are subsets of missing values, outliers, and inconsistent values excluding the empty set. We fix the DQ issue ratio at 20\% to keep the other factors the same except for the DQ issue type.

Due to DQIs, most data show increased scores in the DQ dimension. Furthermore, ML model performance is improved in all data due to DQI, and the p-value of the K-S test is over than 0.05. We observed that the system produces different results based on the ratio and type of DQ issues. These results demonstrate that resolving DQ issues does not automatically lead to improved model performance. The effectiveness of DQI depends not only on the number or type of issues addressed but also on how well the selected methods align with the data’s characteristics and the assumptions of the target model. Although the K-S test confirms distributional changes following DQI, such changes do not consistently yield performance gains across datasets and models. This emphasizes the importance of applying DQI methods in a context-sensitive manner. Instead of using a uniform approach, users are encouraged to explore the visual analytics system to identify appropriate strategies and determine their optimal sequencing based on specific analytical goals.
Further details are available at \url{https://github.com/DVL-Sejong/openlab/tree/main/research/Visual%20Analytics/d-DQIVAR}.
\section{User Study}
This section describes the efficiency and usability of d-DQIVAR by comparing it with two other tools. One is OpenRefine~\cite{openrefine}, which is an open-source tool commonly used for data cleaning and transformation. OpenRefine provides DQ issue position, batch processing of DQ issues, and visualizations of feature distributions. However, ML-related features are not supported by OpenRefine. Therefore, we set the same standards for tools by including the use of libraries like the pycaret in OpenRefine. Another tool is the most recent system~\cite{hong2023how} that adopts a data- and process-driven approach similar to ours. We refer to OpenRefine, the recent system, and d-DQIVAR as \textbf{T1}, \textbf{T2}, and \textbf{T3} in this section, respectively. 

\subsection{Study Design}
Our study employed a 3 (tool) × 3 (data) mixed evaluation design. Participants conducted three exploratory analysis sessions, each with a different tool and data.

\textbf{Data.} We used the Beijing air quality dataset used in Section~\ref{subsec:beijing}, the bike sharing dataset in Section~\ref{sec:evaluation}, and the Boston house price dataset in Section~\ref{subsec:house}.

\textbf{Participants.} We recruited 12 participants (mean age = 27.5 years, four women, eight men), including six with five to ten years of computer science experience and six undergraduate students. All participants have no experience with the data and tools in this study. However, since all participants had experience using data analysis tools such as Excel or Tableau, the basic interactions required for data analysis were familiar to them. Also, all participants are familiar with using libraries for ML. Each participant spent approximately an hour in our study and received a \$15 gift certificate as compensation.

\textbf{Study Protocol.} Before the study, we introduced the participants to the data and provided tool tutorials. We asked participants to “improve DQ for the ML model performance” and “perform given tasks”. The tasks refer to questions about the information that can be grasped during the session, such as the feature name with the most missing values. We chose to assign tasks rather than free analysis to keep participants focused and accountable. The participants were given a maximum of 15 minutes to finish each session.

\textbf{Collected Data.} After each session, the participants were surveyed to evaluate the usability of the tools~(\textbf{T1}, \textbf{T2}, \textbf{T3}). They also provided comments and critiques for the tools.

\subsection{Analysis of Expert Knowledge Supporting}
\begin{figure}[t]
	\centering
	\includegraphics[width=\linewidth]{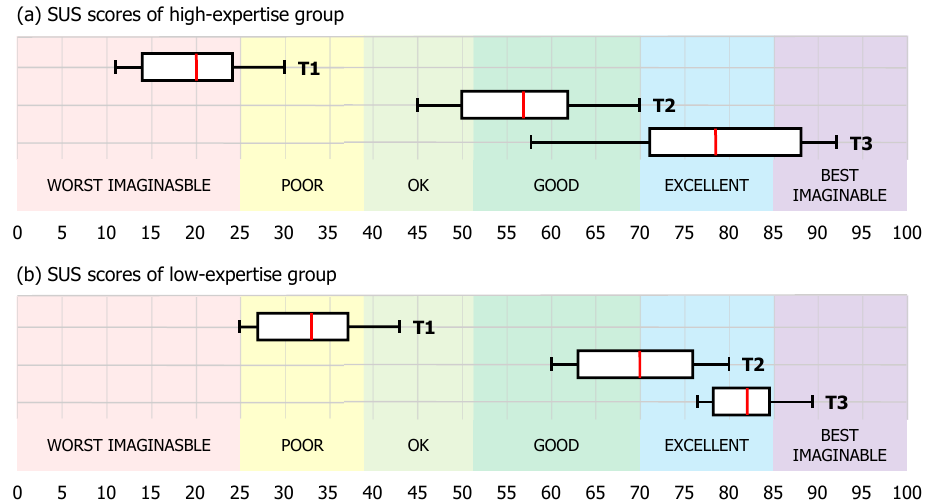}
	\caption{
        SUS scores for OpenRefine~(\textbf{T1}), the most recent system~(\textbf{T2}), and d-DQIVAR~(\textbf{T3}). We calculated the SUS scores of (a) the high-expertise group and (b) the low-expertise group.
	}
	\label{fig:sus}
\end{figure}

To explore the impact of d-DQIVAR in furnishing expertise to less experienced participants, we divided the System Usability Scale~(SUS) scores based on expertise levels as depicted in (a) and (b). Notably, we classified six participants with 5--10 years of experience into a high-expertise group, while six undergraduates were grouped into the low-expertise. The questionnaire consisted of 10 items based on the design considerations presented in Section~\ref{sec:design}. The participants completed the same survey for all sessions and responded on a 1--5 Likert scale to each item. For all questions, the closer the answer is to 1, the more disagree, and the closer to 5, the more agree. Figure~\ref{fig:sus} shows the SUS scores for the questionnaires~\cite{brooke1996sus}.

The average SUS scores of \textbf{T1} are 20 and 33.75, corresponding to \textit{worst imaginable and poor}. The average SUS scores of \textbf{T2} are 56.25 and 69.9, corresponding to \textit{good}. The average SUS scores of \textbf{T3} are 77.5 and 82.5, corresponding to \textit{excellent}. Therefore, the participants rated the usability highly in the order of \textbf{T3}, \textbf{T2}, and \textbf{T1}. In Figure~\ref{fig:sus}~(a), the high-expertise group rated \textbf{T1} as \textit{worst imaginable}, \textbf{T2} as \textit{good}, and \textbf{T3} as \textit{excellent}. Figure~\ref{fig:sus} (b) shows that the low-expertise group assessed \textbf{T1} as \textit{poor}, \textbf{T2} as \textit{good}, and \textbf{T3} as \textit{excellent}. The difference in scores between \textbf{T2} (69.9) and \textbf{T3} (82.6) for the low-expertise group is 12.6, which is less pronounced than the 21-point gap observed between \textbf{T2} (56.25) and \textbf{T3} (77.25) in the high-expertise group. This variation suggests that participants with low expertise are more supported by expert knowledge from the DQI guidance given by our system compared to the high-expertise group.

\subsection{User Feedback}
For \textbf{T1}, the participants mentioned it was difficult to make the DQI for ML because they spent much time on the tasks. In particular, the participants who analyzed the Boston housing dataset considered the original data complete as there were no noticeable DQ issues, such as missing values or inconsistent values in the data. Therefore, the participants who analyzed the Boston housing dataset did not perform a DQI procedure. In addition, the participants responded that it was cumbersome to measure the ML model performance using an external library.

For \textbf{T2}, the participants said that it was easy to identify the optimal DQI but difficult to perform the tasks. Particularly, the participants mentioned that the heatmap, representing the distribution of missing values, outliers, and inconsistent values, was implicitly visualized. Therefore, most of the participants did not perform any additional DQI procedure after selecting the DQI procedure suggested by the system. Also, some participants complained about the delay in the interaction caused by \textbf{T2} calculating all DQI procedures.

The participants responded positively to \textbf{T3} compared to \textbf{T1} and \textbf{T2}. Most participants answered that the detailed analysis of each DQ issue in the DQ assessment view was of great help in performing their work. Some participants noted that the feature that verifies the actual data in the data overview and updates it through DQI is distinct from other tools. Furthermore, some participants reported that they found the feature that displays the change in ML model performance based on DQI and the distortion degree of data characteristics in the data change view to be very useful while confirming the correctness of their actions.

The participants wanted \textbf{T3} to provide additional functions for cleaning data, such as feature filtering or instance modification, like \textbf{T1}. In addition, they wanted \textbf{T3} to provide line charts of ML model performance changes for each DQI stage, just like \textbf{T2}. Participants further suggested improvements for convenience, such as clicking on a donut chart rather than text in the DQ assessment view and increasing the size of the selection and customization buttons in the DQI view. Therefore, we plan to investigate the feedback in the future.
\section{Discussion}

\begin{table}[t]
    \caption{
        Comparison of computing time and MAPE before and after sampling on two real-world datasets.
    }
    \label{tab:scalability}
    \resizebox{\columnwidth}{!}{%
    \begin{tabular}{crrrr}
    \toprule
    Dataset & \multicolumn{1}{c}{\# of records} & \multicolumn{1}{c}{$Time~(\text{sec})$} & \multicolumn{1}{c}{$Time_{s}~(\text{sec})$} & \multicolumn{1}{c}{$\Delta MAPE$} \\
    \midrule
    Stock Market & 851,265 & 1,520,434 & 1,800 & 0.1296 \\
    \midrule
    CalCOFI  & 864,863 & 7,702,099 & 2,267 & 0.1050 \\ 
    \bottomrule
    \end{tabular}}
\end{table}

Identifying optimal DQI procedures using d-DQIVAR is computationally intensive, not because all possible DQI procedures are generated, but because evaluating each candidate requires assessing model performance across 25 machine learning models per dataset to ensure model-agnostic applicability (Section~\ref{subsec:DQIP}). To mitigate this computational burden as the dataset size increases, we integrated a data sampling mechanism. The required sample size is determined using a standard formula for statistical representativeness, which typically yields 385 samples under conventional assumptions~\cite{krejcie1970determining}. To preserve the original data distribution, d-DQIVAR adopts at least 1,000 sampled records, which are validated using the rejection sampling and the Kolmogorov–Smirnov (K–S) test.

To evaluate the effectiveness of sampling, we used two real-world datasets: Stock Market~\cite{stockMarket} and CalCOFI~\cite{calcofikaggle}. The Stock Market dataset comprises 851,265 records of historical daily prices for all tickers traded on NASDAQ. For our experiments, we selected one target feature (volume) and four input features (high, low, close, open). To simulate data quality issues, 30\% of the entries in this dataset were synthetically corrupted. The CalCOFI dataset contains 864,863 records of oceanographic and larval fish observations. From this dataset, we selected one target feature (\textit{T\_degC}) and nine input features (\textit{Depthm}, \textit{Salnty}, \textit{STheta}, \textit{Reclnd}, \textit{R\_Depth}, \textit{R\_TEMP}, \textit{R\_POTEMP}, \textit{R\_SALINITY}, \textit{R\_DYNHT}). Unlike the Stock Market dataset, CalCOFI already contains approximately 15\% of naturally occurring data quality issues. In Table~\ref{tab:scalability}, $Time$ denotes the total computation time on the full dataset, while $Time_s$ represents the time using sampled data. The metric $\Delta \text{MAPE}$ captures the difference in model performance between quality-improved full data and quality-improved sampled data.

Table~\ref{tab:scalability} shows that sampling significantly reduced computation time by 1,528,634 seconds for the Stock Market dataset and 7,630,107 seconds for CalCOFI. Despite this drastic reduction, the differences in MAPE between full and sampled data remained minimal ($\Delta$=0.1296 for Stock Market and $\Delta$=0.1050 for CalCOFI), confirming the effectiveness of sampling-based DQI procedures. These results illustrate a practical trade-off: a negligible loss in predictive accuracy is acceptable given the substantial gains in computational efficiency. This strategy supports scalable and transferable DQI across large and heterogeneous datasets. However, its performance may vary depending on data distribution and task context, which can limit generalizability. As future work, we aim to develop an adaptive framework that dynamically monitors DQI impact and adjusts procedures accordingly~\cite{bayram2024adaptive}.
\section{Conclusion}
In this paper, we proposed a novel data-centric visual analytics system that provides DQI to enhance the performance of ML models. The system computes DQI procedures composed of a data-driven strategy and proposes the optimal DQI procedures to the users. The system addresses data preprocessing techniques such as imputation, duplicate record removal, and feature selection as a data-driven strategy. The system also uses a process-driven strategy to evaluate the DQI and the DQI procedures. The system addresses DQ dimension, ML model performance, and K-S tests as process-driven strategies. The system enhances usability by providing different VA techniques and interactions in the DQ assessment and DQI views. We used the Beijing air quality dataset and the Boston house price dataset to demonstrate how the system improves the DQ. We also showed that the system provides a practical visual analysis process to users. However, since d-DQIVAR focuses on regression, DQ dimension such as class overlap or label purity are not addressed. Therefore, we plan to add DQI methods that are not covered by the system for future extensions. Also, the system only supports DQ for numeric data in CSV format. Therefore, we plan to expand the system to handle data such as images and texts.

\section*{Acknowledgment}
The authors wish to thank A, B, and C. This work was supported in part by a grant from XYZ (\# 12345-67890).

\bibliographystyle{IEEEtran}

\bibliography{0_qualityVA}

\begin{IEEEbiography}[{\includegraphics[width=1in,height=1.25in,clip,keepaspectratio]{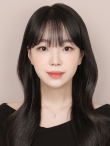}}]{Hyein Hong}
    received the bachelor's and master’s degrees in computer engineering from Sejong University, Seoul, South Korea, in 2021 and 2023, respectively. She is currently a researcher in data visualization lab at Sejong University, Seoul, South Korea. Her research interests include recommendation models, visual analytics, and data quality.
\end{IEEEbiography}

\begin{IEEEbiography}[{\includegraphics[width=1in,height=1.25in,clip,keepaspectratio]{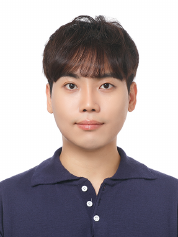}}]{Sangbong Yoo}
    received the bachelor's and Ph.D. degree in computer engineering from Sejong University, Seoul, South Korea, in 2015 and 2022, respectively. He was a postdoctoral researcher at Sejong University, from 2022 to 2025. He is currently a postdoctoral researcher at Korea Institute of Science and Technology (KIST), Seoul, South Korea. His research interests include information visualization, visual analytics, eye-gaze analysis, data quality, and volume rendering.
\end{IEEEbiography}

\begin{IEEEbiography}[{\includegraphics[width=1in,height=1.25in,clip,keepaspectratio]{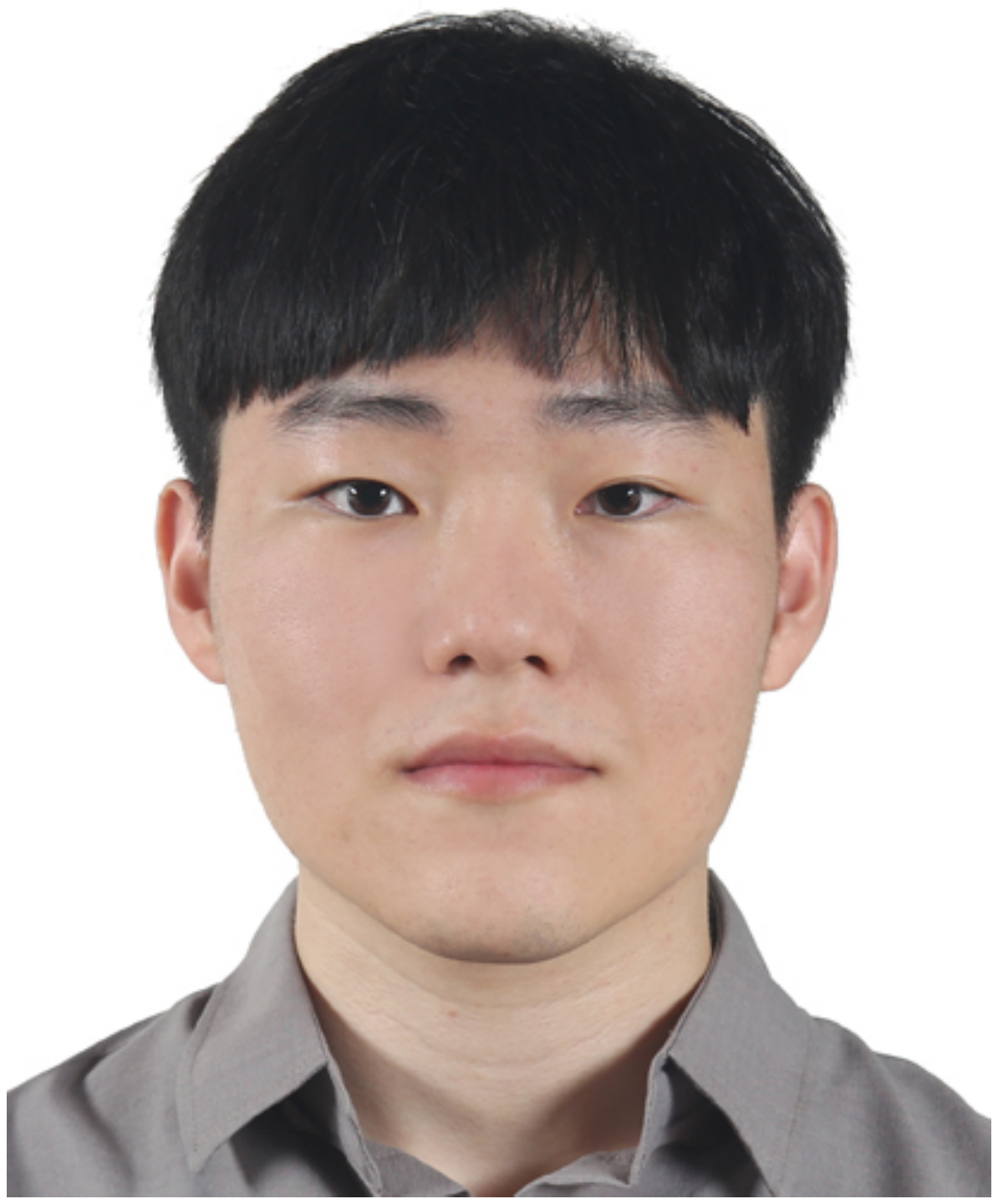}}]{SeokHwan Choi}
    received the bachelor's degree in computer engineering from Sejong University, Seoul, South Korea, in 2022. He is currently a Ph.D. student in the Department of Computer Engineering at Sejong University. His research interests include visual analytics and multi-agent reinforcement learning.
\end{IEEEbiography}

\begin{IEEEbiography}[{\includegraphics[width=1in,height=1.25in,clip,keepaspectratio]{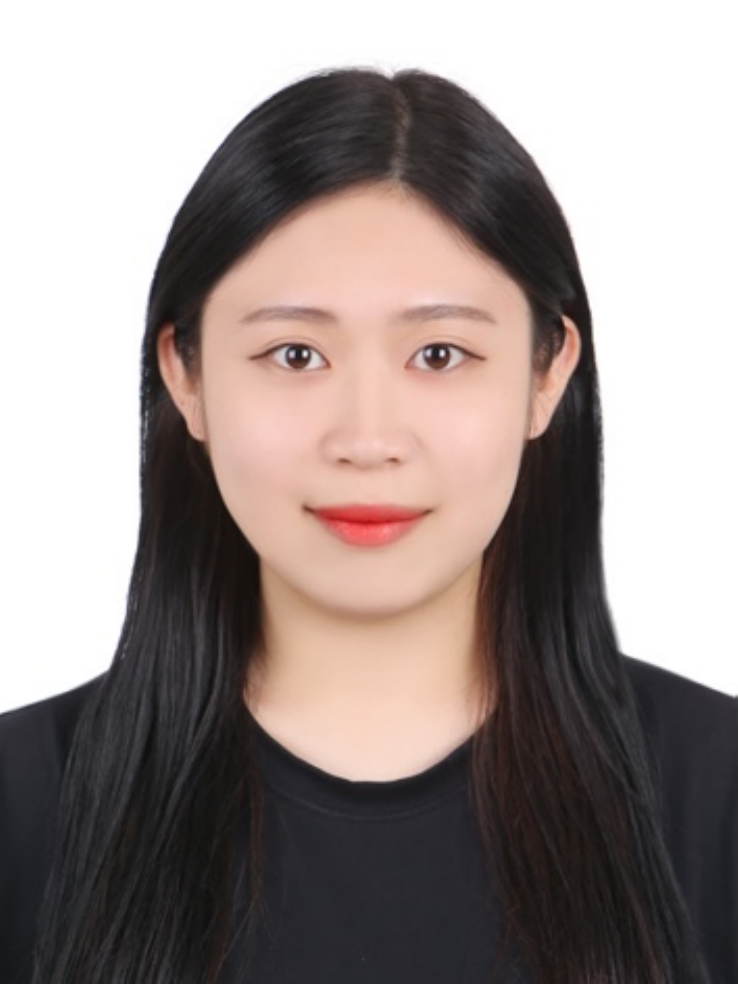}}]{Jisue Kim}
    received the bachelor's degree in computer engineering from Sejong University, Seoul, South Korea, in 2021. She was a developer at APTheFin, from 2022 to 2025. She is currently a developer at Wavebridge, South Korea. Her research interests include back-end programming.
\end{IEEEbiography}

\begin{IEEEbiography}[{\includegraphics[width=1in,height=1.25in,clip,keepaspectratio]{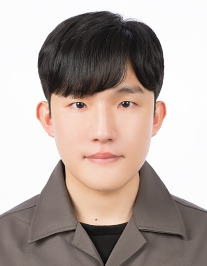}}]{Seongbum Seo}
    received the bachelor's and matser's degrees in computer engineering from Sejong University, Seoul, South Korea, in 2023 and 2025, respectively. He is currently a researcher at Sejong University. His research interests include natural language processing and data visualization.
\end{IEEEbiography}

\begin{IEEEbiography}[{\includegraphics[width=1in,height=1.25in,clip,keepaspectratio]{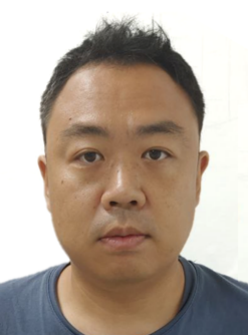}}]{Haneol Cho}
    received the Ph.D. degree in nanomaterial science and engineering from the Korea University of Science and Technology, South Korea, in 2019. Since 2019, he has been a postdoctoral researcher at the Center for Computational Science, Korea Institute of Science and Technology (KIST), focusing on artificial intelligence, large-scale simulation, and optimization for complex systems. In 2025, he joined Sejong University, South Korea, as a research professor. His research interests include AI-driven optimization, computational modeling of social phenomena, and data-driven policy decision-making based on causal inference. His recent work emphasizes applying AI and computational techniques to understand and predict large-scale social dynamics such as disease propagation and collective behavior, with a particular interest in policy-relevant analysis and optimization.
\end{IEEEbiography}

\begin{IEEEbiography}[{\includegraphics[width=1in,height=1.25in,clip,keepaspectratio]{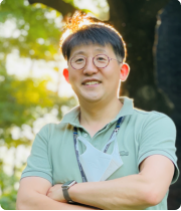}}]{Chansoo Kim}
    is an assistant professor at the University of Science and Technology and a senior research scientist at the Computational Science Centre, Korea Institute of Science and Technology. He leads the AI/R (AI, Information \& Reasoning) Lab., which explores the theoretical foundations of AI, the science of information, and complex (adaptive) systems. His researches span ethics and alignment, optimization, decentralization, and causality in AI—ranging from mathematical theory to real-world applications. Prof. Kim’s work centers on non-Gaussian behaviors—particularly heavy-tailed and leptokurtic—and their applications in learning, inference, finance, and inequality. He traces his academic lineage to C. F. Gau{\ss}. While grounded in theoretical AI, his lab’s research has also informed public policy. During the COVID-19 pandemic, the group supported the Korean CDC and the Office of the President with AI-driven, large-scale agent-based modeling.
\end{IEEEbiography}

\begin{IEEEbiography}[{\includegraphics[width=1in,height=1.25in,clip,keepaspectratio]{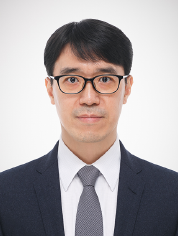}}]{Yun Jang}
    received the bachelor’s degree in electrical engineering from Seoul National University, South Korea, in 2000, and the master’s and Ph.D. degrees in electrical and computer engineering from Purdue University, in 2002 and 2007, respectively. He was a Postdoctoral Researcher at CSCS and ETH Z\"{u}rich, Switzerland, from 2007 to 2011. He is currently a professor in computer engineering at Sejong University, Seoul, South Korea. His research interests include interactive visualization, volume rendering, and visual analytics.
\end{IEEEbiography}

\end{document}